\documentclass[preprint]{elsarticle}

\usepackage{psfrag}
\usepackage{url}
\usepackage{amsmath,amsfonts}
\usepackage{graphicx}

\begin{document}

\title{Statistical Methods for Estimating Complexity from Competition Experiments between Two Populations}

\author[sms]{Stephen J.~Montgomery-Smith}
\ead{stephen@missouri.edu}
\author[fs]{Francis J.~Schmidt\corref{cor}}
\ead{schmidtf@missouri.edu}

\address[sms]{Department of Mathematics, 202 Mathematics Science Building, University of Missouri, Columbia MO 65211,
U.S.A.}
\address[fs]{Department of Biochemistry, 117 Schweitzer Hall, University of Missouri, Columbia MO 65211,
U.S.A.}
\cortext[cor]{Corresponding Author}

\begin{abstract}
Often a screening or selection experiment targets a cell or tissue, which presents many possible molecular targets and identifies a correspondingly large number of ligands. We describe a statistical method to extract an estimate of the complexity or richness of the set of molecular targets from competition experiments between distinguishable ligands, including aptamers derived from combinatorial experiments (SELEX or phage display). In simulations, the nonparametric statistic provides a robust estimate of complexity from a $100 \times 100$ matrix of competition experiments, which is clearly feasible in high-throughput format. The statistic and method are potentially applicable to other ligand binding situations
\end{abstract}


\maketitle

\section{Introduction}

In ecology, the term richness or complexity refers to the number of distinct species that occupy an ecosystem (Hughes et al., 2001; Gotelli and Colwell, 2001). Since the richness of a population in principle can exceed the number of possible observations (i.e., there are more birds than nets, or microorganisms than environmental DNA clones sequences), a common problem is to estimate the number of species not yet observed. A plot of the number of observed species vs. the number of observations is hyperbolic with an asymptote equal to the number of species truly in the population (Hughes et al., 2001; Gotelli and Colwell, 2001). 

The concept of richness or complexity can also apply to molecular systems, as in the use of the term complexity to describe the number of nonreiterated DNA sequences in genomes (Britten and Kohne, 1968). Here we apply the concept to a molecular system, the number of targets that are recognized in screening or combinatorial selection experiments.  

Combinatorial techniques, including phage display and SELEX, start with random libraries and through a reiterative process select ligands capable of binding virtually any target (Tuerk and Gold, 1990; Ellington and Szostak, 1990; Scott and Smith, 1990). The winning molecules from a selection are characterized by strong and specific binding to the species against which they are selected. Nucleic acid and peptide ligands derived from combinatorial selection, termed aptamers, have found diagnostic and therapeutic applications in human and agricultural settings. 

The reiterative selection process in combinatorial selection enriches for species primarily on the basis of their equilibrium binding constants (affinity) for the target. In general, ligands with the greatest affinity are selected most efficiently; however, the affinities of a set of aptamers are usually within an order of magnitude. This phenomenon has been modeled, either directly or by analogy with immune recognition, which like combinatorial technology depends on reiterative selection from an ensemble of potential ligands (Eaton et al. 1995; Lancet et al., 1990). Combinatorial selection schemes have most often been carried out against a single purified molecular target; however, it is also possible to select against complex mixtures of proteins, including whole cells or tissues (Bishop-Hurley et al., 2002; 2005; Zou et al., 2004; Yao et al., 2005). 

It is not always appreciated that the output of a combinatorial selection process contains hidden information about the nature of the target for the selection, analogous to the way in which the number of antibody species after immunization depends on the number of epitopes against which they are selected. A similar process must exist in combinatorial selection where a complex target, e.g., a whole cell or tissue, is used for selection.

Here we develop statistical methods to extend aptamer technology with the goal of estimating the number of distinct binding species in a population of molecular targets, i.e., the complexity or richness of the target population, from the set of ligands that bind to its members. It is important to note that the complexity of this target population is a non-parametric, discrete quantity; i.e., it is the number of distinct species in a population, irrespective of how many individuals are present of each species.

We use as our starting point statistical methods developed by Chao (1984, 1987),
which are used in ecology and population biology to estimate the number of classes
or species in a population, for example, the number of bird species observed in
capture-release experiments, or the number of microbial species found by deep
sequencing DNA from an environmental sample (Hughes et al., 2001). The Chao${}_1$ estimator is given by
\begin{equation*}
S_{\text{Chao}} = S_{\text{observed}} + \frac{r_1^2}{2r_2}
\end{equation*}
where $S_{\text{observed}}$ is the number of species actually counted, $r_1$ is the number observed
only once, and $r_2$ is the number observed exactly twice. We have previously used the Chao$_1$ function to estimate the number of selected peptide aptamer sequences that resulted from a selection experiment (Bishop-Hurley, et al, 2002).

This paper expands statistical estimation to the situation where there are two
populations of ligands, and we can only detect whether a sample
from one population is of the same class or species as a sample from the other
population, i.e., they share a common receptor or target. We show that, if binding of two
combinatorially selected aptamers (or other ligands) can be distinguished from each
other in competition assays, a relatively sparse array of competition assays yields
a robust estimate of the total target population.

\section{Background}

The problem addressed by the Chao$_1$ and related functions is to estimate the
number of unobserved data in an experiment. The original formulation may be
paraphrased as: Suppose we capture 50 birds from an unknown population, and
observe 15 different species. Estimate how many species have not been observed.
While it is clear that one can only get a lower estimate (because there may be a
very large number of extremely rare species which we have essentially no hope of
observing), nevertheless if, say, 4 of the species appear only once, we would assume
that most likely that there are species not observed, because if 15 were the exact
number we would likely have not observed every one of them.

The formula above, derived by Chao (1984, 1987), has found numerous applications in Ecology (Hughes et al., 2001; Chao et al, 2005;  Hughes Martiny et al., 2006). We have used it previously in estimating the population of aptamers from a phage display experiment based on sequences of sampled clones (Bishop-Hurley, et al. 2002). Chao also provided an estimated variance for this estimator. In this paper, we address a variant of this problem.

The application we have in mind is to estimate the complexity of a molecular biological system from binding data only. For example, a problem in developing drugs is to get some idea of the number of functionally diverse surface receptors that are accessible to drug action. This would be useful, for example, in determining if we have sufficiently explored a particular compound series in screening for a new drug. This is of interest because many human disease processes are due to ``undruggable'' targets (Verdine and Walensky, 2009). The question of ``when do we have enough information?'' is inherent in many areas, but hasn't been systematically approached in this aspect of molecular biology.

\section{Mathematical Description}

Let us assume that we have $r$ classes or species, $R_1$, $R_2,\dots,R_r$, of receptors.  The goal is to estimate $r$, the total number of classes of receptor types or classes, defined by the ability to bind different ligands.  The experiment is to select $m$ aptamers from one library and $n$ aptamers from another library.  For example, the two aptamer populations could be selected from two different phage display libraries (such as one on fd filamentous phage, and one on a lytic phage like T7) distinguished by host range. 

A total of $mn$ competition assays are performed, with one aptamer from the first library against another aptamer from the second library.  The results of a small experiment might look something like in Figure~\ref{fig:table} which shows the results of competing $m=11$ first library aptamers against $n=10$ second library aptamers.  An X indicates that the two aptamers compete with each other for a receptor type.  The rows and columns have been rearranged so that the X's form rectangles.  It is clear that in this particular experiment we have observed 5 different receptor types.

For each pair of integers $j,k \ge 0$, we let $r_{j,k}$ denote the number of receptor types that appear in $(j \times k)$ rectangles.  So in the above example, we have $r_{1,1}=1$, $r_{2,2}=2$, $r_{3,2} = 1$, $r_{1,3} = 1$, and all other $r_{j,k}$'s for $i,j\ge 1$ are zero.  Let $r_{j,\bullet} = \sum_{k=1}^\infty r_{j,k}$, and $r_{\bullet,k} = \sum_{j=1}^\infty r_{j,k}$.  So in the example, we have $r_{1,\bullet}=2$.  Let $r_{\bullet,\bullet} = \sum_{j=1}^\infty \sum_{k=1}^\infty r_{j,k}$ be the total number of observed receptor classes.

The quantities $r_{0,0}$, $r_{j,0}$ and $r_{0,k}$ represent the number of receptor classes that bind to none of the displayed aptamers, the number of receptors that bind to $j$ of the first library aptamers and none of the second library aptamers, and the number of receptors that bind to none of the second library aptamers and $k$ of the first library aptamers, respectively.  These numbers cannot be determined by the experiment, and represent the receptors that our experiment failed to detect.

Our goal is to estimate $r = \sum_{j=0}^\infty \sum_{k=0}^\infty r_{j,k}$.  But all we have observed is $r_{\bullet,\bullet} = r - (r_{0,\bullet}+r_{\bullet,0}+r_{0,0})$.  Thus the goal reduces to finding a good estimate for $r_{0,\bullet}+r_{\bullet,0}+r_{0,0}$.

In this paper we propose a lower estimate of the number of classes of receptors by
$$ \hat r = r_{\bullet,\bullet} + \frac{(r_{1,\bullet}+r_{\bullet,1}+r_{1,1})^2}{2r_{2,\bullet}+2r_{\bullet,2}+4r_{2,2}} ,$$
with estimated variance of $\hat r$
$$ \text{Var}(\hat r) = (r_{1,\bullet}+r_{\bullet,1}+r_{1,1})(\tfrac12\rho + \rho^2 + \tfrac14\rho^3) ,$$
where $\rho = (r_{1,\bullet}+r_{\bullet,1}+r_{1,1})/(r_{2,\bullet}+r_{\bullet,2}+2r_{2,2})$, and we interpret $0/0$ as zero.

\section{Mathematical Justification of the Formulae}

For each receptor type $R_i$, denote by $p_i$ the probability that a randomly selected first library aptamer will bind to it, and by $q_i$ the probability that a randomly selected second library aptamer will bind to it.  Note that $\sum_{i=1}^r p_i = \sum_{i=1}^r q_i = 1$.

If probability that the receptor $R_i$ will bind to exactly $j$ of the $m$ randomly selected first library aptamers, and to exactly $j$ of the $n$ randomly selected second library aptamers, is given by the Binomial distribution, and well approximated by the Poisson distribution, as
$$ \binom mj p_i^j (1-p_i)^{m-j} \binom nk q_i^k (1-q_i)^{n-k}
   \approx \frac1{j!k!} (m p_i)^j (n q_i)^k e^{-mp_i-nq_i} .$$
Therefore,
\begin{gather*}
\mathbb E r_{j,k} \approx \frac1{j!k!} \sum_{i=1}^r (m p_i)^j (n q_i)^k e^{-mp_i-nq_i}, \\
\mathbb E r_{j,\bullet} \approx \frac1{j!} \sum_{i=1}^r (m p_i)^j e^{-mp_i}(1-e^{-nq_i}), \\
\mathbb E r_{\bullet,k} \approx \frac1{k!} \sum_{i=1}^r (n q_i)^k e^{-nq_i}(1-e^{-mp_i}).
\end{gather*}
Then the quantity
$$ f(k) = k! r_{k,\bullet} + k! r_{\bullet,k} + (k!)^2 r_{k,k} $$
satisfies
$$ \mathbb E f(k) = \int \phi^k \, d\mu ,$$
where $\phi$ is a positive function on an appropriate measure space.  Then it follows from H\"older's inequality (see for example Royden 1988) that $\log\mathbb E f(k)$ is a convex function.  In particular, $\log\mathbb E f(1) \le \frac12(\log\mathbb E f(0) + \log\mathbb E f(2))$, that is, $\mathbb E f(0) \ge \mathbb E f(1)^2/\mathbb E f(2) $.  Hence we obtain $f(1)^2/f(2)$ as a good lower estimator for $f(0)$.

The justification for the variance formula is as follows.  It seems reasonable to suppose that $P_0=f(0)$, $P_1=f(1)$ and $P_2=f(2)/2$ are independent Poisson random variables.  For example, $r_{2,\bullet}$, $r_{\bullet,2}$ and $r_{2,2}$ are close to being Poisson random variables.  If we assume that $2 r_{2,2}$ is a Poisson random variable, then admittedly the variance of this one variable changes by a factor of two, but quite likely this will be a small error in the large calculation.  Adding $r_{2,\bullet}$, $r_{\bullet,2}$ and $2r_{2,2}$ will not be exactly a Poisson random variable, but again, the distortion won't be excessive.

Next, if $P$ is a Poisson random variable whose expected value is $\lambda$, then it can be shown that $\mathbb E(P^n I_{P>0}) = \lambda^n(1+n(n-1)/2\lambda+O(\lambda^{-2}))$ as $\lambda\to\infty$.

Now, since $\hat r = r - f(0) + f(1)^2/f(2) = r-P_0+\frac12 P_1^2/P_2 $, we can suppose that $\text{Var}(\hat r) \approx \text{Var}\, P_0 + \frac14\text{Var}(P_1^2/P_2)$.  We also suppose that $\mathbb E f(k) \approx f(k)$.  Thus
\begin{align*}
\text{Var}(P_1^2/P_2)
&\approx
\mathbb E P_1^4 \mathbb E I_{P_2>0} P_2^{-2} - \big( \mathbb E P_1^2 \mathbb E I_{f(2)>0} P_2^{-1} \big)^2 \\
&\approx P_1^4 P_2^{-2} \left(1+\frac6{P_1}\right)\left(1+\frac3{P_2}\right) 
 - P_1^4 P_2^{-2} \left(1+\frac1{P_1}\right)^2\left(1+\frac1{P_2}\right)^2 \\
&\approx 6P_1^3 P_2^{-2} + 3P_1^4 P_2^{-3} - 2 P_1^3 P_2^{-2} - 2 P_1^4 P_2^{-3} .
\end{align*}
Furthermore, $\text{Var}\, P_0 \approx P_0 \approx \frac12 P_1^2/P_2$.  Therefore
$$ {Var}(\hat r) \approx P_1 (\tfrac12 \rho + \rho^2 + \tfrac14 \rho^3) ,$$
where $\rho = P_1/P_2$.

\section{Numerical Simulation}

\begin{figure}
\begin{center}
\includegraphics{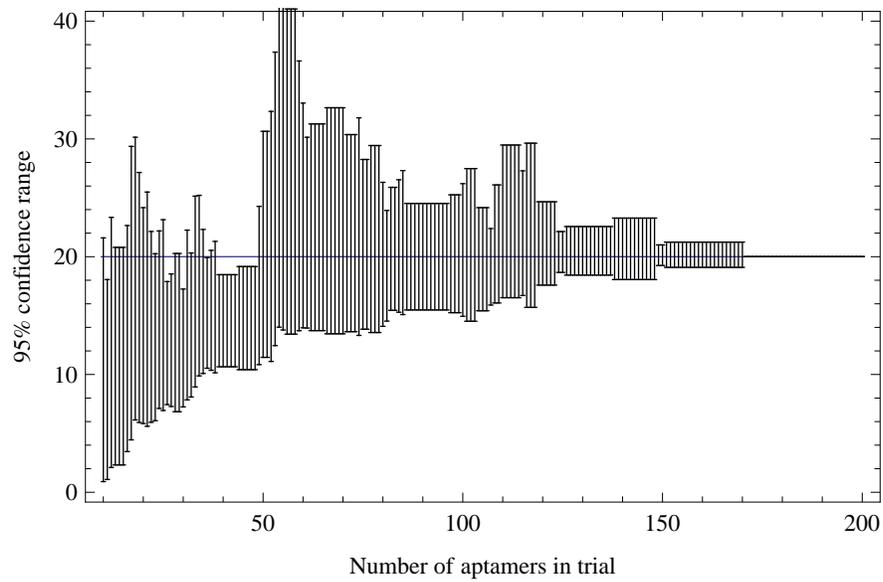}
\caption{95\% confidence range for receptome size as function of
number of aptamers in trial.}
\label{fig:graph}
\end{center}
\end{figure}

In the Figure~\ref{fig:graph}, we show the results of this statistic using simulated data with receptome size equal to 20.  The horizontal axis shows the number of aptamers from the fist library, which we assume to be the same as the number of aptamers from the second library.  The vertical axis represents the estimated receptome size, with the vertical lines representing the 95\% confidence intervals.

Even though this is only simulated data, it looks as if the estimator converges reasonably quickly as the sample size becomes larger.

\section{Discussion}

The analysis presented here suggests a readily accessible means to determine the
richness of a molecular population from competitive binding data. In simulations,
a sparse sample of binding aptamers resulted in an estimation that converged on
the ``true'' number of cellular receptors, converging on a confidence limit within
20\% of the real value when a grid of $100 \times 100$ aptamers are evaluated.
In ecology, the complexity or richness of a system refers to the number of biological
species in a community (Gotelli and Colwell, 2001; Hughes et al., 2001). Members of a biological species are defined by the inability to produce fertile offspring. Here we define the richness of a molecular community by the ability to be recognized by individual ligands: if two molecules are recognized by the same ligand or aptamer population they are regarded as members of the same species.

The richness of a molecular population viewed in this manner is potentially larger than the richness of the same population estimated from bioinformatics data. Protein products of variant splicing from the same genetic locus could be recognized by the same aptamers in a population, in which case they would be regarded as members of the same species (Kim and Lee, 2008). Conversely, they would be regarded as different species if there were aptamers that recognized one variant rather than another. Currently, alternative splicing variants are predicted by sequence comparison of EST or other data. The data are generally filtered to remove transcripts that do not correspond to known splice sites, a process that can result in false negatives (Kim and Lee, 2008). The techniques described here could provide a complementary estimate of diversity since the structures of alternative splicing variants would likely be recognized by different aptamers (Scott and Smith, 1990; Smith and Petrenko, 1997).

Competition of aptamers is not an all-or-none phenomenon; rather, it depends on the relative affinities and concentrations of the competing aptamers. One might expect that the analysis described here would be complicated by differing affinities of the competing aptamers. For example, if the $K_d$ values of two aptamers were to differ by several orders of magnitude, then the higher-affinity aptamer might not be displaced by the lower-affinity one if the latter concentration were only tenfold higher than the former. In the reciprocal experiment, a lower-affinity aptamer would be displaced efficiently by a higher-affinity one. This non-reciprocal competition would confound the experiment envisioned in Figure~\ref{fig:table}. This difficulty is not likely to be serious, however. Both theory and experiment indicate that the affinities of the aptamers resulting from selection experiments are distributed fairly narrowly (Eaton et al., 1995; Lancet et al., 1990). Thus, the population of winning aptamers would have roughly equivalent affinities, and a large fraction would mutually compete in binding assays.

The binding algorithm described here may have other uses. For example, high throughput screening of large compound libraries it is often employed to find compounds that affect growth or differentiation of cultured cells. If the compounds can be distinguished by a chemical signal (e.g., by differential labeling), the number of targets that the compounds recognize could be estimated from relatively few experiments.

\section{Conflict of Interest Statement}

None declared

\section{Acknowledgments}

We thank members of the MU Combinatorial Biotechnology group for discussions. This work was supported by the University of Missouri Research Board and by National Aeronautics and Space Agency grant NNX07AJ21G.

\vfill\eject

\begin{figure}
\begin{center}
\begin{tabular}{|c|c|c|c|c|c|c|c|c|c|c|c|}
\cline{2-12}
\multicolumn{1}{c}{}&\multicolumn{11}{|c|}{first library}\\
\multicolumn{1}{c}{}&\multicolumn{11}{|c|}{aptamers}\\
\hline
             & x &   &   &   &   &   &   &   &   &\phantom{x}   & \phantom{x}  \\
\cline{2-12}
             &   & x & x & x &   &   &   &   &   &   &   \\
\cline{2-12}
             &   & x & x & x &   &   &   &   &   &   &   \\
\cline{2-12}
second library &   &   &   &   & x & x &   &   &   &   &   \\
\cline{2-12}
aptamers     &   &   &   &   & x & x &   &   &   &   &   \\
\cline{2-12}
             &   &   &   &   &   &   & x & x &   &   &   \\
\cline{2-12}
             &   &   &   &   &   &   & x & x &   &   &   \\
\cline{2-12}
             &   &   &   &   &   &   &   &   & x &   &   \\
\cline{2-12}
             &   &   &   &   &   &   &   &   & x &   &   \\
\cline{2-12}
             &   &   &   &   &   &   &   &   & x &   &   \\
\hline
\end{tabular}
\end{center}
\caption{Example results of competition between aptamers from the two libraries.}
\label{fig:table}
\end{figure}

\vfill\eject
\
\vfill\eject

\vfill\eject

\end{document}